\documentstyle[aps,epsf,rotate,preprint]{revtex}

\begin{document}
\newcommand{\be}{\begin{equation}}
\newcommand{\ee}{\end{equation}}
\newcommand{\bea}{\begin{eqnarray}}
\newcommand{\eea}{\end{eqnarray}}


\title{Power-law Distribution of Family Names in Japanese Societies}

\author{
Sasuke Miyazima$^1$, Youngki Lee$^2$, Tomomasa Nagamine$^1$
and Hiroaki Miyajima$^3$
}

\address{
$^1$Department of Engineering Physics, Chubu University,
      Kasugai, Aichi 487-8501, Japan \\
$^2$Center for Polymer Studies \& Department of Physics,
      Boston University, Boston, MA 02215, USA \\
$^3$Department of Space Science, Ohio State University, Columbus,
      OH 43210, USA
}

\date{Last modified: December 1, 1998; Printed: \today}

\maketitle

\begin{abstract}

We study the frequency distribution of family names.
From a common data base,
we count the number of people who share the same
family name. This is the size of the family.
We find that (i) the total number of different family names in a society
scales as a power-law of the population,
(ii) the total number of family names of the same size decreases as
the size increases with a power-law and
(iii) the relation between size and rank of a family name also shows
a power-law. These scaling properties are found to be consistent for five
different regional communities in Japan.

\end{abstract}

\bigskip

Scaling laws have been playing an important role in science for the past
several decades \cite{widom}.  Diverse systems in nature have been found
to exhibit a scaling law and self-similarity without a fine tuning of
external parameters ---known as self-organized-criticality.  A simple
model proposed by Bak {\it et.~al.} \cite{bak} shows that the minimal
ingredients of these scaling behaviours is to have a large number of
degrees of freedom and nonlinear interactions between them.  Human
societies also show complexity which meets the above features of
self-organized criticality.  In this context, many of human activities
including word freqeuncy \cite{zipf65}, traffic flow \cite{gerlough},
economics \cite{econophysics}, population growth \cite{population}, city
growth \cite{allen}, internet \cite{internet}, citation frequency
\cite{citation} and war distribution \cite{war} have been reported to
show scaling behaviour.

Here, we study frequency distribution of family names
in Japanese societies.
We define a {\it family} as a group of people who share the same
family name, i.e., the different families are identified by their
family names.
We also define the {\it size of a family}, $s$, as a number of people
in that family. We rank families by their size from the biggest family
to the smallest family;
For example, the biggest family in town of Fuso is
``Senda'' with family size $s(Senda)=296$, so its rank is $r(Senda)=1$.
The second biggest is ``Kondo'' with size $s(Kondo)=229$,
and rank $r(Kondo)=2$, and so on  \cite{comment_b}.
In this way we measure the size and the rank of all families.

\begin{figure}[htb]
\centerline{
        \epsfxsize=10.0cm
        \rotate[r]{\epsfbox{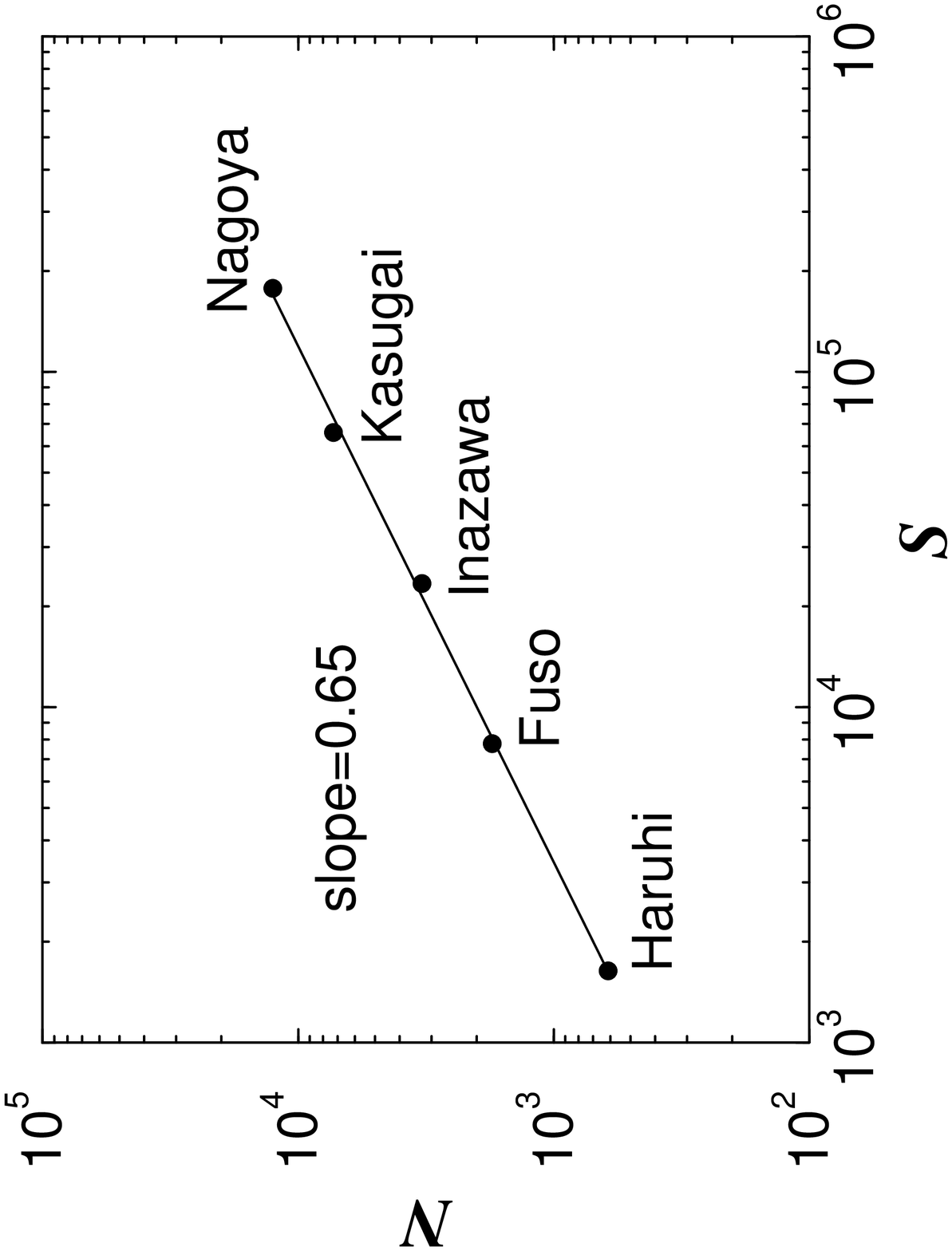}}
}
\vspace*{1.0cm}
\caption{The number of family names $N$ against the total population
$S$ for five different regional societies in Japan shows power-law
behaviour as $N \sim S^\chi$.
By linear regression in double logarithmic plot we estimate
the exponent ${\chi}=0.65 \pm 0.03$.
}
\label{fig:1}
\end{figure}

We analyze the telephone directories of five regional communities
in Japan: town of Haruhi, town of Fuso, city of Inazawa,
city of Kasugai and 1/3 of the city of Nagoya.
The directories were published in 1998 by the communications company
``NTT". The total number of customers $S$ appeared in these directories
are $1634$, $7775$, $23365$, $65988$ and $177267$, respectively.
First, we count the number of different family names, $N$, appeared
in the directories. In Fig.~\ref{fig:1} we plot $N$ versus $S$
and we find that
\be
N \sim S^{\chi},
\label{eq:chi}
\ee
with an exponent $\chi=0.65 \pm 0.03$.

Next, we investigate the scaling properties of two different quantities:
(i) the distribution of the family size $n(s)$
which is the number of families of the same size $s$,
and (ii) the relation between size and rank of a family,
i.e., $s(r)$ which forms the so-called Zipf's plot \cite{zipf65}.
The two quantities are complementary in a sense
that $n(s)$ mainly focuses on the scaling property of the smaller size family
while $s(r)$ highlights the scaling property of the bigger size family.
We find the power-law scalings for both the quantities which are consistent for all five regions investigated.

\begin{figure}[htb]
\centerline{
        \epsfxsize=7.0cm
        \rotate[r]{\epsfbox{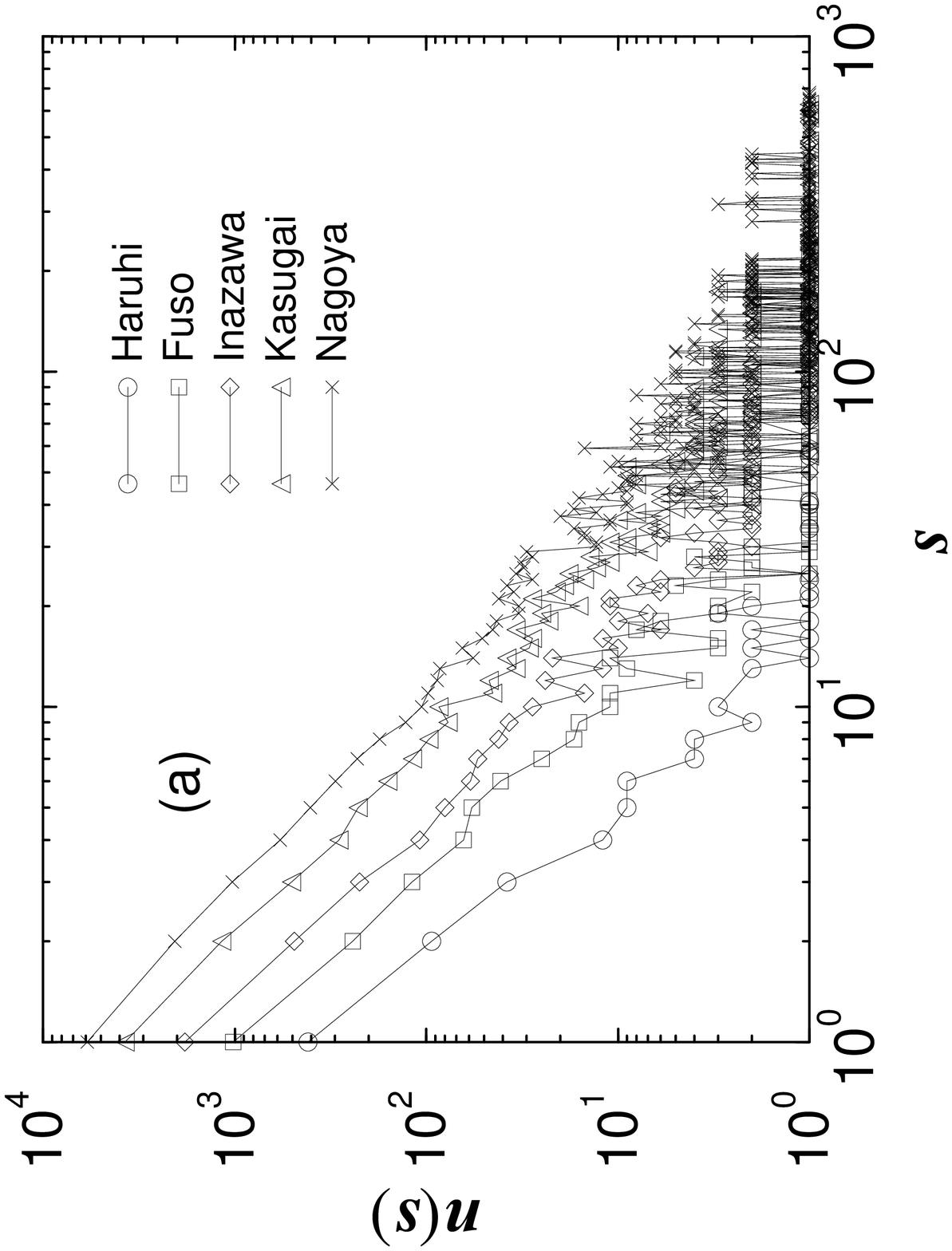}}
}
\vspace*{1.0cm}
\centerline{
        \epsfxsize=7.0cm
        \rotate[r]{\epsfbox{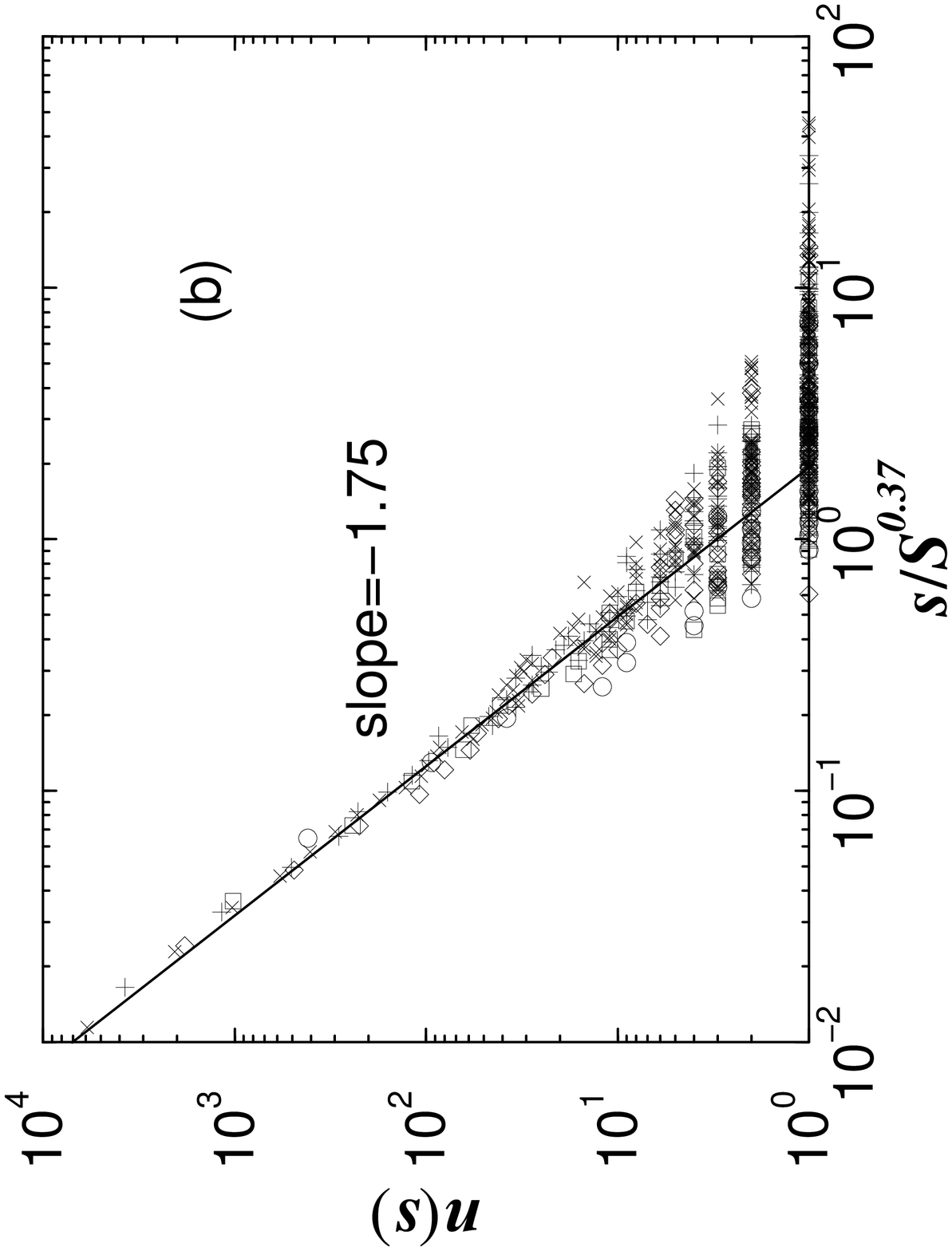}}
}
\vspace*{1.0cm}
\caption{
a) The double logarithmic plot of the histogram $n(s)$ vesus family size $s$
for five regions in Japan.
b) Data collapse using the scaling form in Eq.~(\protect{\ref{eq:scaling1}}).
The linear fit of the power-regime gives $\tau=1.75 \pm 0.05$.
}
\label{fig:2}
\end{figure}

We measure the distribution $n(s)$ for each town
which is shown in Fig.~\ref{fig:2}a in double logarithmic
scale. It shows a nice power-law behaviour with same
exponent for all five different communities.
We suggest the following scaling form for $n(s)$ \cite{stauffer}:
\be
n(s)=Af({s \over s^*}),
\label{eq:scaling1}
\ee
where the scaling function $f(x)$ behaves as $f \sim x^{-\tau}$
for $x \ll 1$ and $f = 1$ for $x \gg 1$.
Here $s^*$ is a characteristic family size at which
$n$ becomes one, i.e. $n(s^*)=1$, which in turn gives $A=1$.
In Fig.~\ref{fig:2}b we try to collapse data
using the scaling form of Eq.~(\ref{eq:scaling1})
with an additional scaling law $s^* \sim S^\alpha$
and the scaling exponent $\alpha=0.37 \pm 0.03$.
A linear fit of the collapsed scaling function yields
$\tau=1.75 \pm 0.05$.

From the normalization condition,
\be
\int_1^{s_*} n(s) ds =N,
\label{eq:norm1}
\ee
and the scaling form for $n(s)$ [Eq.~(\ref{eq:scaling1})]
we obtain a relation, $s^* \sim N^{1/\tau}$.
This scaling, combined with our finding $N \sim S^\chi$,
gives
\be
\alpha={\chi \over \tau}.
\label{eq:scaling2}
\ee
This scaling relation is well consistent with the exponents
measured within error bars.

\begin{figure}[htb]
\centerline{
        \epsfxsize=7.0cm
        \rotate[r]{\epsfbox{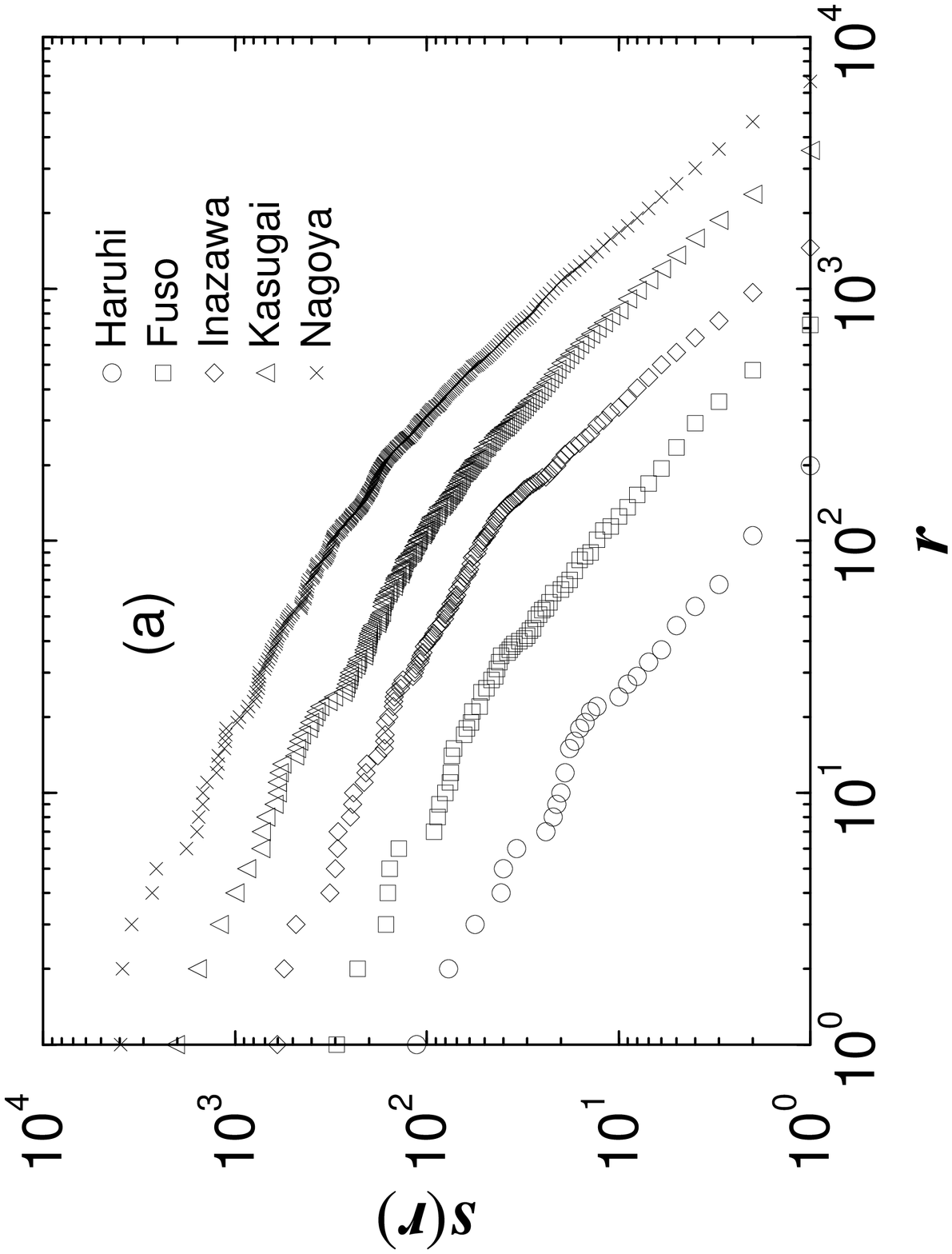}}
}
\vspace*{1.0cm}
\centerline{
        \epsfxsize=7.0cm
        \rotate[r]{\epsfbox{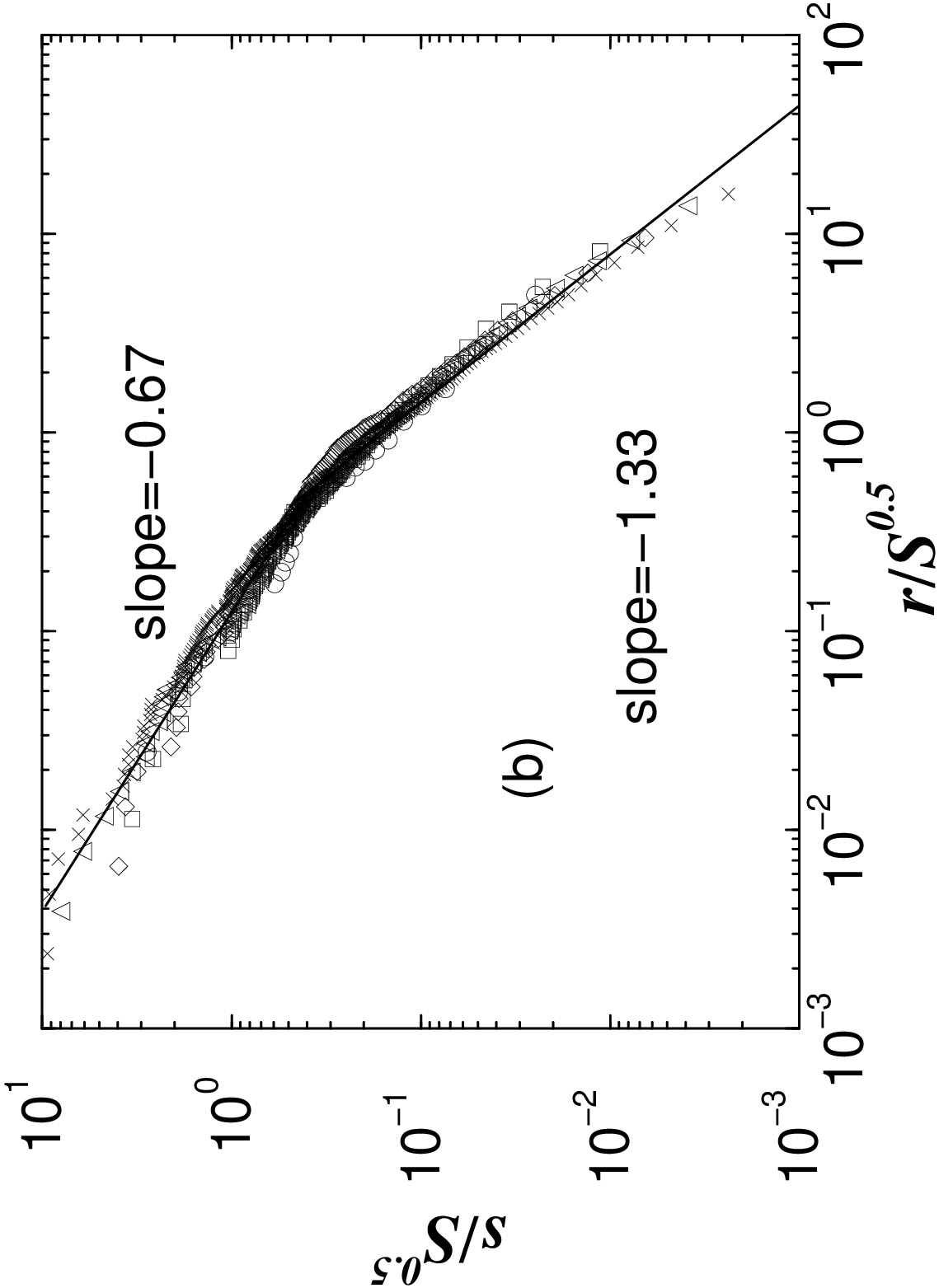}}
}
\vspace*{1.0cm}
\caption{
a) The size of a family name $s(r)$ when plotted against
the rank of the family name $r$ shows a crossover behaviour
at the characteristic rank $r^*$ where $r^*$ scales as
$S^{\alpha'}$. The solid line connects the crossover points
whose slope is one implying $s(r^*) \sim r^*$.
b) Data collapse using the scaling form in Eq.~(\protect{\ref{eq:scaling2}}).
A crossover behaviour is observed from the $\phi_I=0.67 \pm 0.03$ to $\phi_{II}=1.33 
\pm 0.03$.
}
\label{fig:3}
\end{figure}

In Fig.~\ref{fig:3}a we plot the family size $s$ versus rank $r$
in double logarithmic scale.
Each curve shows a crossover behaviour from one power-law regime
with exponent $\phi_I=0.67 \pm 0.03$, to another steeper power-law
decay with exponent $\phi_{II}=1.33 \pm 0.03$ at the characteristic
rank $r^*$ which also scales as $r^* \sim S^{\alpha'}$.
We propose the following scaling form for $s(r)$;
\be
s(r)=r^*g({r \over r^*}),
\label{eq:scaling3}
\ee
where the scaling function $g$ behaves as $g \sim x^{-\phi_I}$
for $x \ll 1$ and $g \sim x^{-\phi_{II}}$ for $x \gg 1$.
In Fig.~\ref{fig:3}b we try to collapse the data using
the scaling form of Eq.~(\ref{eq:scaling3}) and
the best fit is obtained when $\alpha'=0.5 \pm 0.05$.

Two quantities, $n(s)$ and $r(s)$, are related by an integral equation \cite{cohen97};
\be
r(s) = \int_s^\infty n(s')ds' \sim s^{1-\tau}.
\ee
By inverting the relation we obtain a scaling relation,
$s(r) \sim r^{ 1 \over 1-\tau}$.
This relation gives the exponent
\be
\phi_{II}={ 1 \over \tau-1},
\label{eq:exponent1}
\ee
because the scaling exponent $\tau$ is measured for small $s$, i.e.
for $r>r^*$.
Note that the Eq.~(\ref{eq:exponent1}) is well satisfied
by our results.
The fact that the crossover points $r^*$ scales as $S^{0.5}$
suggests that the sampling of the population is random so that relative
deviation of the probability decreases as $S^{-0.5}$ as number of data
points $S$ increases. 

\begin{figure}[htb]
\centerline{
        \epsfxsize=5.0cm
        \rotate[r]{\epsfbox{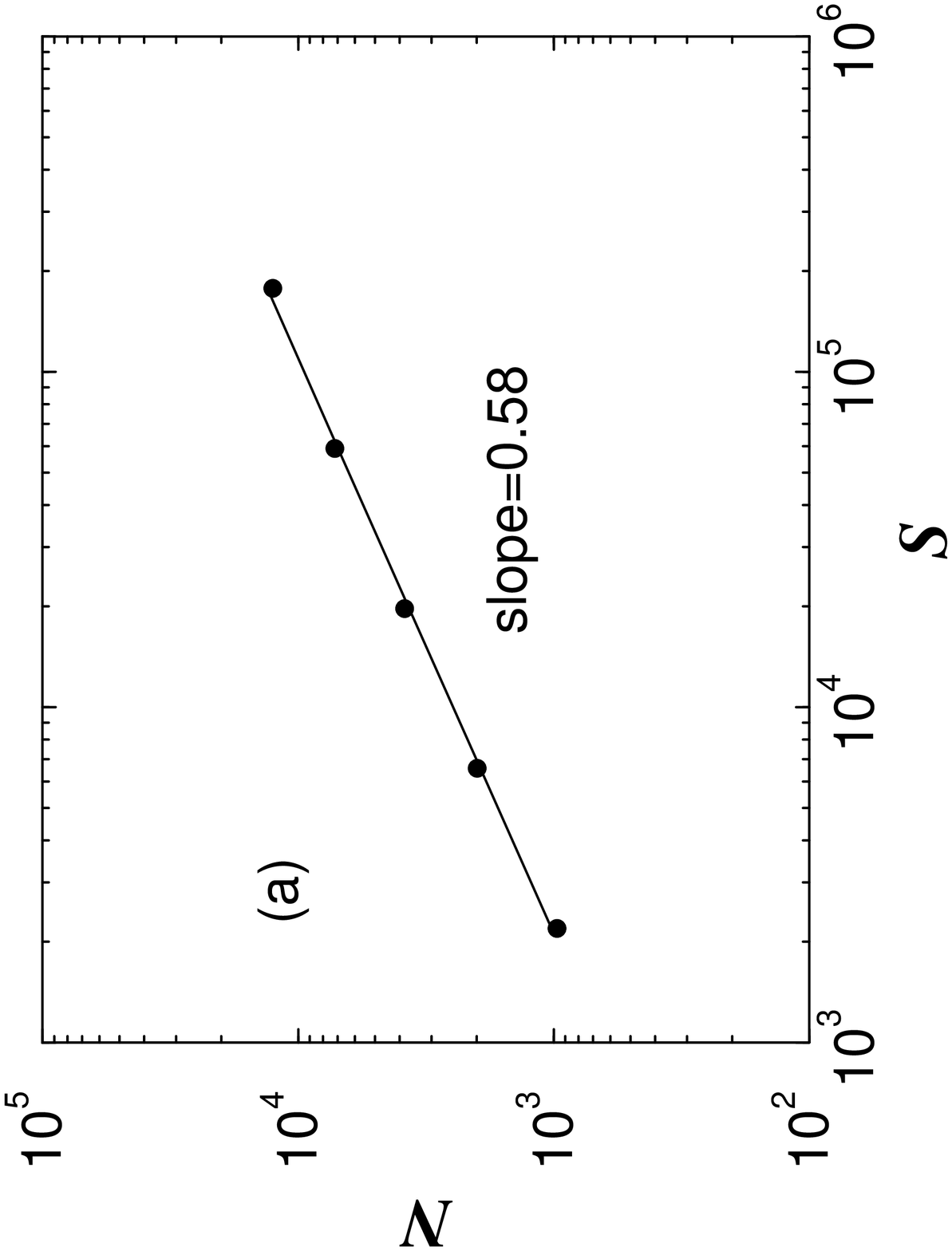}}
}
\vspace*{1.0cm}
\centerline{
        \epsfxsize=5.0cm
        \rotate[r]{\epsfbox{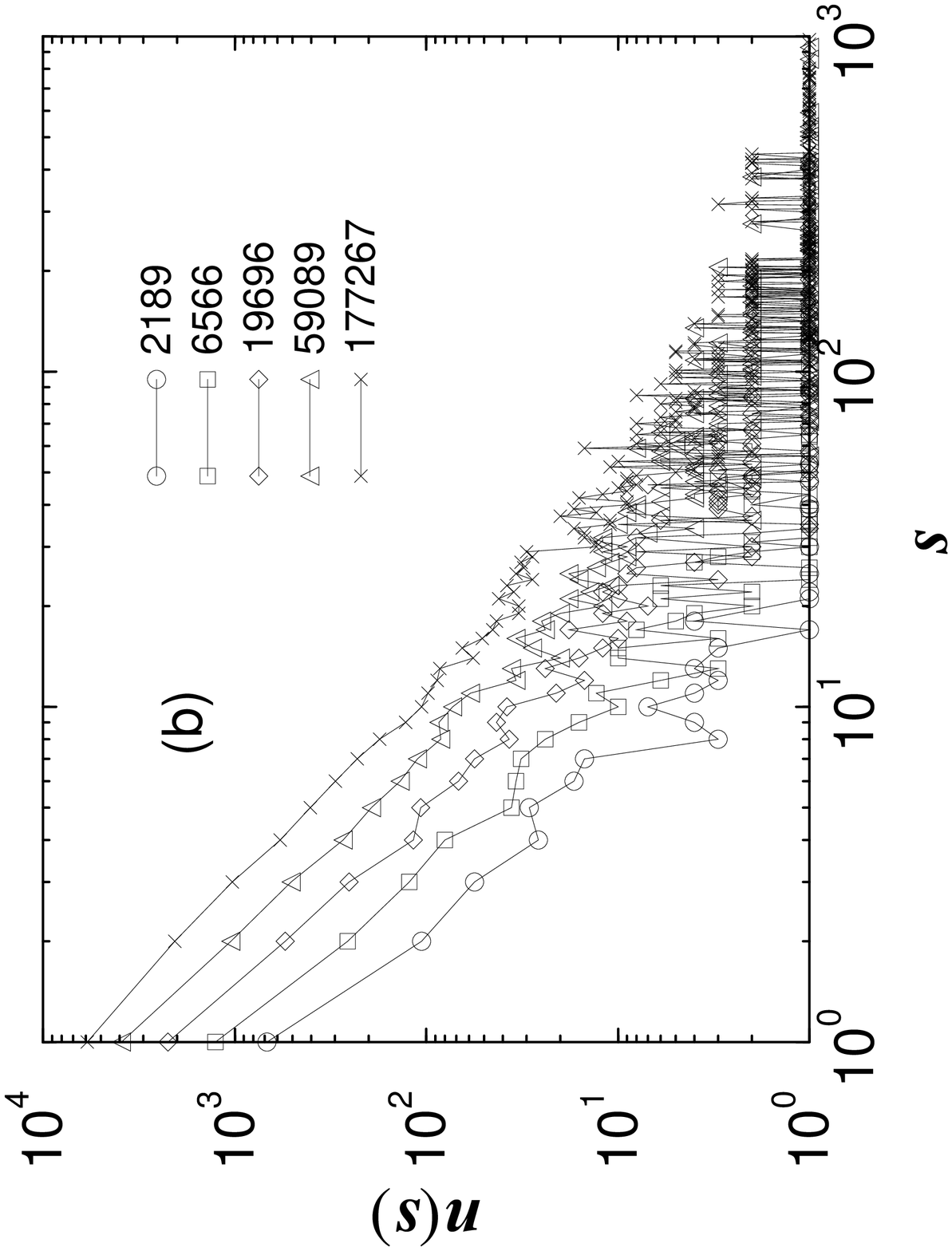}}
}
\vspace*{1.0cm}
\centerline{
        \epsfxsize=5.0cm
        \rotate[r]{\epsfbox{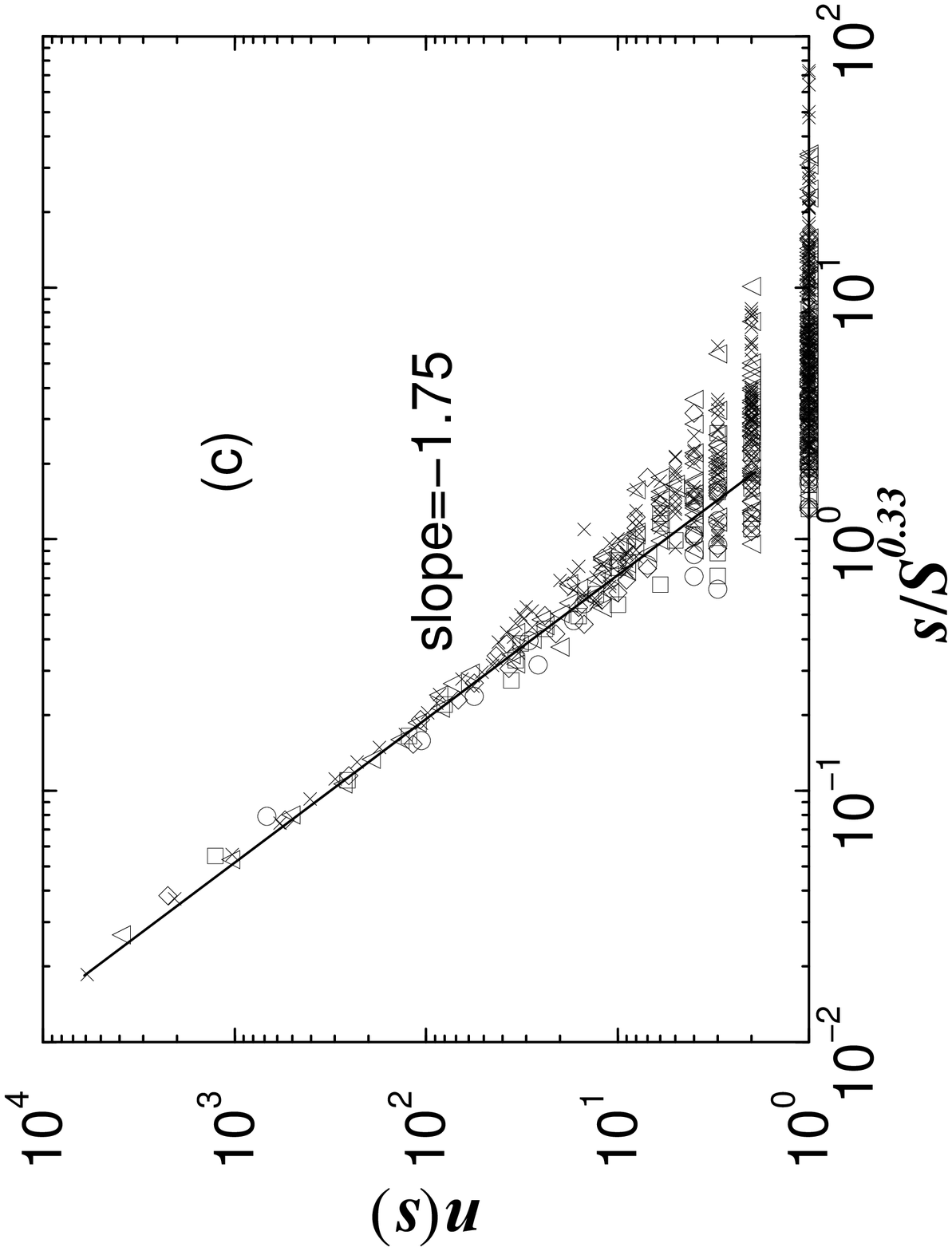}}
}
\vspace*{1.0cm}
\caption{
a) The double logarithmic plot of $N$ versus $S$ obtained from the distributions for the randomly chosen population $S=2189, 6566, 19696, 59089$ and $177267$.
It shows a simple power-law relation, $N \sim S^\chi$ with $\chi=0.58$.
b) The double logarithmic plot of the number of family names of same size $s$, $n(s)$, versus the size $s$.
c) Data collapse using the scaling form in Eq.~(\protect{\ref{eq:scaling1}}).
The linear fit of the power-regime gives $\tau=1.75$.
}
\label{fig:4}
\end{figure}

To test the role of the communities on the observed scaling behaviours
we randomly select a population and repeat our analysis for the
extracted data set.
Figure \ref{fig:4} shows the distributions for the randomly chosen population
$S=2189, 6566, 19696, 59089$ and $177267$ out of the biggest data for $1/3$ of city of Nagaya.
It shows very close scaling behaviours as Figs.~\ref{fig:1} to
\ref{fig:3}.
This experiment suggests that {\it the families are distributed randomly
in the town without spatial correlation.}
Such scaling universality in the family structure of contemporary
societies could be explained as a result that the
time scale characterizing the migration of pupulation in a community
is much shorter than the time scale asscoiated with the reproduction 
of a famiy name.


The scaling exponents $\tau=1.75$, $\phi_{I}=0.67$ and $\phi_{II}=1.33$
are different from the Zipf's result on word frequency where the
exponents are $\tau=2.0$ and $\phi=1.0$.   
The power-law relation between $N$ and $S$ and it's exponent
$\chi=0.65$ observed in family name distribution seem to be nontrivial.
One may expect this scaling law breaks if the number of available family
names in a society is too small compared to the population.
Cohen {\it et.~al.} \cite{cohen97} found that this situation occurred in the
words frequency distribution --- for very large $S$, $N(S)$ approaches a
plateau. They found that the exponent $\chi$ for the number of different
words in a text is also a function of length of the text.
This is true also for the societies where the family names are strictly
inherited from fathers to sons without any creation of new family names.
In fact, the expectation number of sons per parents is one under
the stationary constant population.
Then the survival probability  $P(t)$ of a family name after
$t$ generations decreases as $P(t) \sim t^{-0.5}$.
As a result, after many generations, only a few family names
will dominate the whole population in the society.
This is the situation in countries
where the creation of new family names has been strictly restricted
for many generations such as in Korea.
The total number of family names in Korea is about $250$ while
the total population is about $50$ millions.
On the contrary Japan has most rich family names in the world
whose total number of family names is about $132,000$ and
the population is about $125$ millions.
The creation of a new family name in Japan is also very rare.
However, historically the most of Japanese family names were created
about $120$ years ago \cite{history}.
The short history of family names may cause to preserve
the diversity and the scaling properties of family names
as it was at the creation.

In summary, we have investigated the distribution of Japanese family names
for five different regional communities in Japan. From the our
empirical investigation, the power-law relation between total number of
different family names and total population appeared in a telephone
directory with the exponent $\chi = 0.65$. Also we have
found that the name-variety-size distribution shows nice power-law
scaling with the exponent $\tau = 1.75$ and the cutoff exponent, $\alpha =
0.37$. These scaling properties are consistent
for five regional communities and randomly generated societies with
with different populations.
In a size-rank distribution of family name we have obtained
a crossover behaviour from one exponent, $\phi_I = 0.67$ to another
exponent $\phi_{II} = 1.33$ at the crossover point $r^* \sim S^{\alpha'}$
with $\alpha '= 0.5$. This result is consistent
even if the specific family names of higher rank in one community is
different from those in other communities.
We have also derived scaling relations between these
exponents.

\bigskip
Acknowdgements

\medskip
We thank I. Grosse, P.Ch. Ivanov and S. Havlin for helpful discussions.

\newpage

\newpage

\newpage

\newpage


\begin{thebibliography}{99}
\bibitem{comment_b}
Note that if two or more family names are of same size,
then we assign the same number for their rank and skip
the next available numbers by the number of degeneracy
and assign the new number for the rank of the next biggest
family.
\end{thebibliography}
\end{document}